\documentstyle[11pt,aaspp4]{article}

\received{}
\accepted{}

\def\chisqnu{$\chi^2_\nu$}
\def\RXTE{{\it RXTE}}
\def\CGRO{{\it CGRO}}

\slugcomment{Submitted to ApJ.
Refereed version.
Draft: 981204}

\lefthead{Smith et al.}
\righthead{Multiwavelength Observations of GX 339--4}

\begin{document}

\title{Multiwavelength Observations of GX 339--4 in 1996.\\
I. Daily Light Curves and X-ray and Gamma-Ray Spectroscopy}

\author{I. A. Smith\altaffilmark{1}, 
E. P. Liang\altaffilmark{1},
D. Lin\altaffilmark{1},
M. Moss\altaffilmark{1,2},
A. Crider\altaffilmark{1},
R. P. Fender\altaffilmark{3},
Ph. Durouchoux\altaffilmark{4},
S. Corbel\altaffilmark{4},
R. Sood\altaffilmark{5}}

\altaffiltext{1}{Department of Space Physics and Astronomy, 
Rice University, MS 108, 6100 South Main Street, Houston, TX 77005-1892}
\altaffiltext{2}{Currently at McDonnell-Douglas, St. Louis, MO}
\altaffiltext{3}{Astronomical Institute `Anton Pannekoek', Center for 
High-Energy Astrophysics, University of Amsterdam, Kruislaan 403, 
1098 SJ Amsterdam, The Netherlands}
\altaffiltext{4}{DAPNIA, Service d'Astrophysique, CE Saclay,
91191 Gif sur Yvette Cedex, France}
\altaffiltext{5}{School of Physics, ADFA, Northcott Drive, Canberra ACT 2600,
Australia}

\begin{abstract}
As part of our multiwavelength campaign of GX 339--4 observations in 1996
we present our radio, X-ray, and gamma-ray observations made in July, when
the source was in a hard state (= soft X-ray low state).

The radio observations were made at the time when there was a possible 
radio jet.
We show that the radio spectrum was flat and significantly variable, and
that the radio spectral shape and amplitude at this time were not anomalous 
for this source.
Daily light curves from our pointed observation July 9--23 using the Oriented 
Scintillation Spectrometer Experiment (OSSE) on the {\it Compton 
Gamma-Ray Observatory} (\CGRO),
from the Burst and Transient Source Experiment (BATSE) on \CGRO, and
from the All Sky Monitor on the {\it Rossi X-Ray Timing Explorer} (\RXTE) 
also show that there was no significant change in the X- and gamma-ray
flux or hardness during the time the possible radio jet-like feature was seen.

The higher energy portion of our pointed \RXTE\ observation made July 26 
can be equally well fit using simple power law times exponential (PLE) 
and Sunyaev-Titarchuk (ST) functions.
An additional soft component is required, as well as a broad emission
feature centered on $\sim 6.4$ keV.
This may be an iron line that is broadened by orbital Doppler motions and/or
scattering off a hot medium.
Its equivalent width is $\sim 600$ eV.
Our simplistic continuum fitting does not require an extra reflection 
component.
Both a PLE and a ST model also fit our OSSE spectrum on its own.
Although the observations are not quite simultaneous, combining the \RXTE\ 
and \CGRO\ spectra we find that the PLE model easily fits the joint spectrum.
However, the ST model 
drops off too rapidly with increasing energies to give an acceptable joint fit.
\end{abstract}

\keywords{binaries: spectroscopic --- stars individual (GX 339--4) ---
black hole physics --- X-rays: stars --- accretion: accretion disks}

\section{Introduction and Overview}

Most Galactic black hole candidates exhibit at least two distinct spectral
states (see Liang \& Narayan 1997, Liang 1998, Poutanen 1998 for reviews).
In the hard state (= soft X-ray low state) the spectrum from $\sim$ keV 
to a few hundred keV is a hard power law (photon index $1.5 \pm 0.5$) with 
an exponential cutoff.
This can be interpreted as inverse Comptonization of soft photons.
In the soft state (often, but not always, accompanied by the soft X-ray 
high state), the spectrum above $\sim 10$ keV is a steep power law 
(photon index $> 2.2$) with no detectable cutoff out to $\sim$ MeV.
This multi-state behavior is seen in both persistent sources 
(e.g. Cygnus X-1) and transient black hole X-ray novae (e.g. GRS 1009--45).
GX 339--4 is unusual in that it is a persistent source, being detected by 
X-ray telescopes most of the time, but it also has nova-like flaring states.

In 1996, we performed a series of multiwavelength observations of GX 339--4
when it was in a hard state (= soft X-ray low state).
This paper is one of a series that describes the results of that campaign.
In Paper II (\cite{smi99II}) we discuss the rapid X-ray timing variability in 
an observation made by the {\it Rossi X-Ray Timing Explorer} (\RXTE) 
1996 July 26 UT.
In Paper III (\cite{smi99III}) we discuss our Keck spectroscopy performed on 
1996 May 12 UT.
These papers expand significantly on our preliminary analyses 
(\cite{smi97a,smi97b}).

We start this paper by presenting in \S 2 the radio spectra from 1996 July 
taken when there was a possible radio jet (\cite{fen97}).
We show that the radio spectrum is flat and significantly variable,
and that the spectral shape and amplitude were not unusual for the source
at that time.

In \S 3 we discuss the observing and data analysis details for our 
pointed observation 1996 July 9-23 using the Oriented Scintillation 
Spectrometer Experiment (OSSE) on the {\it Compton Gamma-Ray Observatory}
(\CGRO).

In \S 4 we show the daily X-ray and gamma-ray light curves from 1996 July.
In addition to our OSSE results, we show the data obtained by the All Sky 
Monitor (ASM) on \RXTE\ and by the Burst and Transient Source Experiment 
(BATSE) on \CGRO\ using the Earth occultation technique.
The soft X-ray flux during our observations was very low, and the
spectrum was very hard.
However, unlike at the time of our Keck observation on 1996 May 12, the
source was significantly detected during most of the month.
Most importantly, there was no significant change in the higher 
energy emission during the time the radio jet-like feature was seen.

In \S 5, we present the OSSE spectrum and fit it with simple power law 
times exponential (PLE) and Sunyaev-Titarchuk (ST; \cite{suny80}) functions.
They both give equally good fits to the OSSE data alone.

In \S 6, we discuss the observing and data analysis details for the
1996 July 26 \RXTE\ observation.

In \S 7, we present the X-ray spectroscopy from the \RXTE\ observation.
Both a PLE and a ST model can explain the \RXTE\ data alone.
An additional soft component is required, as well as a {\it broad} emission
feature centered on $\sim 6.4$ keV.
This may be an iron line that is broadened by orbital Doppler motions and/or 
scattering off a hot medium.
Its equivalent width is $\sim 600$ eV.

Combining the \RXTE\ and \CGRO\ spectra, we show in \S 8 that the PLE model 
easily fits the joint spectrum, but that the ST model does not.
In B\"ottcher, Liang, \& Smith (1998) we use the GX 339--4 spectral data 
to test our detailed self-consistent accretion disk corona models.

\section{Radio Observations}

The radio counterpart to GX 339--4 was discovered in 1994 by the
Molonglo Radio Observatory, Australia (MOST) at 843 MHz (\cite{sood94}).
Monitoring over the following years has found that the radio emission
is variable, with a flux density $\lesssim 10$ mJy 
(\cite{sood97,cor97,han98}).

High resolution 3.5 cm radio observations with the Australia Telescope 
Compact Array (ATCA) detected a possible jet-like feature 
1996 July 11--13 (MJD 50275--7; \cite{fen97}).
An ATCA observation 1997 Feb 3 may have a small extension in the 
direction opposite this jet, but no strongly significant confirmation
of the jets has been found (\cite{cor97}).

Figure 1 shows the daily radio spectra in 1996 July.
All the data points are from ATCA, except the one at 843 MHz from MOST
(\cite{han98}).
The ATCA observations are the ones made at the time that the possible jet-like
feature was detected: see Fender et al. (1997) for the observing details.

The radio spectrum is approximately flat, and shows a significant variability.
{\it The spectral shape and amplitude were not anomalous for this
source during the time of the possible radio jet}. 

The radio emission suggests the continual presence of an outflow in this state.
The approximately flat spectrum is not consistent with optically
thin emission (unless the electron distribution is exceptionally hard)
and is indicative of some absorption in the radio-emitting regions. 
One possible geometry would be a conical jet such as discussed in 
Hjellming \& Johnston (1988).

\section{OSSE Observations and Reductions}

OSSE observed GX 339--4 as a ToO during an outburst for 1 week beginning 
1991 September 5.
The OSSE flux was $\sim 300$ mCrab between 50 and 400 keV (\cite{grab95}).
A second 1 week observation was carried out beginning 1991 November 7,
when the source was $\sim 40$ times weaker (\cite{grab95}).
Our observation was made 1996 July 9-23 (MJD 50273--287).
The flux is a factor of $\sim 2$ lower than that found in the (brighter)
1991 September observation.

The OSSE instrument consists of four separate, nearly identical NaI-CsI
phoswich detectors with a field of view $3.8 \times 11.4$ degrees FWHM
(see Johnson et al. (1993) and Grabelsky et al. (1995) for 
instrumental details and observing techniques).
Our observations consist of a series of alternating on- and off-source
pointings with 2.048 minutes spent on each pointing.
The on-source pointings (at $l = 339.0\arcdeg$, $b=-3.7\arcdeg$)
were centered close to GX 339--4 ($l = 338.9\arcdeg$, $b=-4.3\arcdeg$), 
with the long dimension of the collimator oriented perpendicular to the 
Galactic plane.
The total on-source observing time (per detector) was $1.59 \times 10^5$
seconds.
The off-source pointings were located along the galactic plane at 
$\pm 12\arcdeg$ from the on-source pointing (they were centered at
$l = 351.0\arcdeg$, $b=-3.2\arcdeg$ and $l = 327.0\arcdeg$, $b=-4.1\arcdeg$).
Both backgrounds gave the same results, indicating there were no bright
sources in either one.

Version 7.4 of the IGORE OSSE data analysis package was used to 
subtract the background and to generate the daily light curves and spectra 
in the standard way (\cite{john93}).
To perform joint fits with the \RXTE\ spectrum, we used IGORE to generate
a spectrum and response matrix that could be read by XSPEC.

Grabelsky et al. (1995) estimated the contribution from the diffuse
component from the Galactic plane.
They found that this is a negligible effect when the source is bright,
as in our observation.
The spectrum of the diffuse component also has a similar shape to that
of GX 339--4.
We have therefore not subtracted an estimate for the diffuse component.

We checked all of our results for the individual detectors separately,
and found no discrepancies.
We have therefore combined all four detectors in all the results 
presented here.

\section{Daily Light Curves}

Figure 2 shows the daily light curves for our OSSE data in the 50--70 and 
70--270 keV bands, as well as the hardness ratio of these two bands.
These are combined with the data from the \RXTE\ ASM and BATSE
Earth occultation.
The BATSE light curve assumes an optically thin thermal
bremsstrahlung (OTTB) model with a fixed $kT = 60$ keV (\cite{rubin98}).

The soft X-ray flux in 1996 July was low, and the spectrum was very hard.
However, unlike at the time of our Keck observation on 1996 May 12, the
source was significantly detected during most of the month.
During the month, the fluxes at all energies were generally rising.
Figure 1 of Smith et al. (1997a) and Figure 1 of Rubin et al. (1998)
show that the X-ray and gamma-ray fluxes peaked on $\sim$ TJD 50290.

A constant does not fit the OSSE 50--70 keV light curve.
The reduced \chisqnu = 3.86 for $\nu = 14$ degrees of freedom:
the probability that a random set of data points would give a value of
\chisqnu\ as large or larger than this is $Q = 1.3 \times 10^{-6}$.
Similarly, a constant does not fit the OSSE 70--270 keV light curve
with \chisqnu = 2.88, $\nu = 14$, $Q = 2.2 \times 10^{-4}$.
For both bands, the linear relationships shown in Figure 2 give good fits
($Q = 0.35, 0.44$ respectively).

While there is an indication from the OSSE (70--270 keV)/(50--70 keV)
hardness ratio that the spectrum may be softening during the two week 
observation, this is not statistically significant.
A constant fit to the hardness ratio gives acceptable results,
\chisqnu = 1.11, $\nu = 14$, $Q = 0.34$.
This fact and the linear rise in the flux are important in \S 8, when we 
combine the non-simultaneous \RXTE\ and OSSE spectra.

Most importantly, {\it there was no significant change in the higher 
energy emission during the time the possible radio jet-like feature was seen}.
One might have expected that the physics behind the jet formation could 
have led to a significant change in the higher energy emission.
But since the radio flux was not unusually bright during this time, it
is possible that the energy release in this case was relatively small.
Further multiwavelength observations during radio jet or radio flaring 
events will be very important to understanding whether the synchrotron
emitting electrons are involved in the Compton scattering that produces
the hard X-rays.
In other black hole candidates, violent changes in the high energy emission 
may or may not be associated with large radio flares, e.g. GRO J1655--40
(\cite{har95,tav96}).

\section{OSSE Spectra}

The OSSE spectrum was extracted by averaging over the whole two week
observation.
Since the hardness ratio did not change significantly over this time,
this gives a reliable measure of the spectral shape during these
two weeks, though Figure 2 shows that the normalization on any given
day will be different from the average presented here.

The OSSE spectrum was fit over the 0.05--10 MeV energy range using a
forward folding technique.
Concerns about the OSSE instrumental response and the precision of the 
cross calibration of the detectors (\cite{grab95}) are mitigated by 
looking at the results for the individual detectors separately.
We found that the results were consistent in all four detectors, and thus
we present the results for the four detectors combined.

Throughout this paper we will only consider simple phenomenological models
for the spectral fitting: see B\"ottcher, Liang, \& Smith (1998) for detailed
self-consistent accretion disk corona model fits to all the GX 339--4 data.
In the OSSE range, the spectrum has a power law shape with a cut-off.
We found that it is only necessary to use one model component to explain the 
OSSE spectrum, and we consider two simple functional forms:

\noindent
1) Power law times exponential (PLE).
In this model, the flux has the form
$F(E) \propto E^{-\alpha} \exp(-E/kT)$.
An optically thin Compton spectrum can be roughly described using this
model (\cite{haa93}).

\noindent
(2) Sunyaev-Titarchuk function (ST; \cite{suny80}).
In this thermal Comptonization model with a spherical geometry, for energies 
below $kT$ of the scattering medium, the spectrum is a power law with photon 
index $\alpha = -(1/2) + [(9/4) + \gamma]^{1/2}$, where
$\gamma = \pi^2 m c^2 / 3 k T (\tau + 2/3)^2$, 
and $\tau$ is the optical depth.

Figure 3 shows the best PLE and ST model fits to the OSSE data. 
Both functional forms give equally good fits {\it to the OSSE data alone}.
See \S 8 for the effect of doing joint fits with the \RXTE\ spectrum.

The best fit PLE model has $\alpha = 1.15 \pm 0.07$, $k T = 97 \pm 6$ keV, and 
flux $0.094 \pm 0.001~{\rm photons}~{\rm cm}^{-2}{\rm s}^{-1}{\rm MeV}^{-1}$
at 100 keV.
(The errors are $1 \sigma$).
This has \chisqnu = 1.01, $Q = 0.37$.
The flux normalization at 100 keV is a factor of 2.0 lower than that found
by Grabelsky et al. (1995) in the 1991 September observation, while 
our power law index is slightly steeper (theirs was $0.88 \pm 0.05$), 
and our $k T$ is higher (they had $68 \pm 2$).

The best fit ST model has $\tau = 2.5 \pm 0.1$, $k T = 47 \pm 1$ keV, and
flux $0.093 \pm 0.001 ~{\rm photons}~{\rm cm}^{-2}{\rm s}^{-1}{\rm MeV}^{-1}$
at 100 keV.
This has \chisqnu = 1.00, $Q = 0.45$.
The flux normalization at 100 keV is a factor of 2.0 lower than that found
by Grabelsky et al. (1995) in the 1991 September observation, while 
our $\tau$ is smaller (theirs was $3.0 \pm 0.1$), and 
our $k T$ is higher (they had $37 \pm 1$).
However, the value we derive for $\alpha = 1.9$ is the same as theirs.

Grabelsky et al. (1995) noted that their ST fit was marginal at best.
They found that the model dropped off much more rapidly than the data at
higher energies.
As we have shown, this disagrees with our results where we get good fits 
to the OSSE data alone.
{\it However, in \S 8 we will show that this is no longer the case when the
OSSE data is combined with the \RXTE\ spectrum.}

\section{\RXTE\ Observations and Reductions}

The \RXTE\ pointed observation was made as a Target of Opportunity on
1996 July 26 (MJD 50290), just after the OSSE run ended.
We generated the \RXTE\ spectrum using two of its instruments, the Proportional 
Counter Array (PCA), and the High Energy X-ray Timing Experiment (HEXTE).

The PCA consists of five Xe proportional counters with a total effective 
area of about 6500 cm$^2$ (\cite{jah96}).
These cover a scientifically validated energy range of 2 -- 60 keV with 
an energy resolution $< 18$\% at 6 keV.

The HEXTE consists of two independent clusters each containing four 
phoswich scintillation detectors (\cite{roth98}).
These have a total effective area of about 1600 cm$^2$, and cover an 
energy range of 15 -- 250 keV with an energy resolution of 15\% at 60 keV.
Each cluster can ``rock'' (beam switch) along mutually orthogonal directions 
to provide background measurements.
In our observation, the background offset was $1.5\arcdeg$ with a switching 
time of 16 seconds.

Both instruments have a $\sim 1\arcdeg$ field of view.
No other X-ray sources were in the GX 339--4 field of view, or in the
HEXTE background regions.
For both instruments, the background dominates at higher energies.
For GX 339--4 this is at $\gtrsim 70$ keV for the PCA and $\gtrsim 130$ keV 
for HEXTE.
The PCA does not make separate background measurements.
Instead we used version 1.5 of the background estimator program.
The HEXTE background was determined from the off source pointings in
both chopping directions.
For both instruments, we found that beyond the energies where the background
dominated, the model (PCA) and observed (HEXTE) background measurements were
the same as what was observed in the on-source observation.
This gives us confidence that the background measurements are valid.
The backgrounds have been subtracted from all the spectra shown here.

For both instruments, data were only used when the spacecraft was not 
passing through the South Atlantic Anomaly and when the source was observed
at elevations $> 10\arcdeg$ above the Earth's limb and the pointing
was offset $< 0.02\arcdeg$ from the source.
The observation was short, with good data starting 18:20:32.625 and 
ending 20:14:48.497 UT.
Because of the above constraints, valid data was only available in four
segments: see Smith \& Liang (1999) for the light curve.
{\it We combined all this data to make one spectrum, although it should be
cautioned that the source was extremely variable during this observation}
(\cite{smi99II}).
For the PCA, the total on-source exposure time was 3424 sec.
After correcting for dead time, this gave an effective exposure of 3337 sec.
The dead time correction was more important for the HEXTE data, where the
time taken to rock also has to be accounted for in the background exposures.
For HEXTE, the adjusted on-source exposure time was 1085 sec for cluster A,
and 1070 sec for cluster B.

For the PCA, we used the Standard 2 production data set to generate the
spectrum with 129 spectral channels.
We used the \RXTE\ tasks in FTOOLS version 4.0 to extract the data.
We performed the extraction twice, once using all five Proportional
Counter Units (PCU), and once using just the top PCU layer.
The top layer is responsible for $\sim 80$\% of the detected counts,
so has the best S/N.
Using all the layers gives a higher count rate, and thus a better sensitivity
at higher energies, but with a lower S/N.
The background contributes $\sim 13$\% of the on-source count rate when using
all the layers, and $\sim 8$\% using just the top layer.
We found that the spectral results made little difference which of these two
methods was used, and we chose to retain both of them for our fitting.
For both cases, we used version 2.1.2 of the PCA response matrices.
We ignored the PCA data below 2 keV that are not scientifically valid.
We also ignored the PCA data above 60 keV, well below where the background
dominated: we found that making this upper limit to the energy range smaller
had no effect on the spectral fitting.
We simply dropped the data around 4.78 keV where there is a problem with the 
response matrix due to the xenon L edge.

For the HEXTE, we used the E\_8us\_256\_DX1F event list data.
These record each individual event with $8 \mu$sec timing and
256 energy channel resolution.
As with the PCA, the \RXTE\ tasks in FTOOLS version 4.0 were used to 
extract the two HEXTE spectra, one for each cluster.
For both clusters, we used the 1997 March 20 versions of the HEXTE 
response matrices.
We ignored the HEXTE data above 120 keV, well below where the background
dominated.
We found that changing this upper limit to the energy range 
had no effect on the spectral fitting.
Given the short exposure time, the HEXTE error bars are large, and the 
spectra do not extend far enough beyond the break energy to be constraining.
We also ignored the HEXTE data below 20 keV.
Again, this choice of energy cut-off had no effect on the fitting results.

We used the two PCA and two HEXTE spectra jointly in our spectral fitting.
This was performed using XSPEC version 10.0.
The relative overall normalization of the PCA and HEXTE spectra was left as a
free parameter in the fitting, while all the other model parameters were the 
same for both instruments.
We found that the relative normalization was $0.71 \pm 0.01$ in all of
the results presented here.
We did not find it necessary to include separate overall normalizations 
between the two PCA spectra or between the two HEXTE spectra.
The fitting was performed using the complete spectral resolution of both
instruments, though the data have been rebinned in the figures so that 
each bin is at least $5 \sigma$.

\section{\RXTE\ Spectra}

As in \S 5, in this paper we only consider two simplistic phenomenological
models for the higher energy emission.
A detailed fitting of this data using a more self-consistent model
is presented elsewhere (\cite{bls98}).

\subsection{Power Law Times Exponential Fit}

A PLE model fits the \RXTE\ data (alone) above 15 keV.
The cut-off energy is very poorly determined by the \RXTE\ data 
because the HEXTE error bars are large in this short observation, and
the \RXTE\ data does not sample far beyond the cut-off energy.
We therefore chose to fix $kT = 96.9$ keV at the value found for the 
best OSSE fit.

Figure 4 shows the results of fitting the PLE model to the data above 15 keV.
This uses $\alpha = 1.26 \pm 0.01$, and flux 
$0.130 \pm 0.005~{\rm photons}~{\rm cm}^{-2}{\rm s}^{-1}{\rm keV}^{-1}$
at 1 keV.
(The errors are 90\% confidence regions for varying one parameter).
This has \chisqnu = 0.93, $Q = 0.79$ (for fitting above 15 keV).

It is apparent from Figure 4 that two extra components are required to 
fit the spectrum at lower energies:

\noindent
(1) A soft component that peaks at $\sim 2.5$ keV.
Such a soft component has been seen in previous GX 339--4 observations,
eg. by {\it Tenma} (\cite{mak86}), {\it EXOSAT} (\cite{ilo86,men97}), 
and {\it Ginga} (\cite{miy91,ueda94}).
It is present in all states, and dominates during the soft state.

\noindent
(2) A {\it broad} emission centered on $\sim 6.4$ keV.
A similar feature was seen in {\it Ginga} observations of GX 339--4 in the 
low (=hard) state, and was modeled using a reflection model (\cite{ueda94}).
Using a broad iron line improved the fits in the {\it EXOSAT} observations
of GX 339--4 made in the high (=soft) state (\cite{ilo86}).
A broad line was seen in \RXTE-{\it ASCA} observations of Cygnus X-1
in the soft (=high) state (\cite{cui98}).
A very similar effect was seen in \RXTE\ observations of Cygnus X-1 in the
low (=hard) state by Dove et al. (1998), though they chose to force the
Fe K$\alpha$ line to be narrow (with a line width of 100 eV giving an
equivalent width of 60 eV) which resulted in obvious residuals.
\RXTE\ observations of the Seyfert galaxy MCG --5-23-16 also found this
feature, confirming previous {\it ASCA} observations (\cite{wea98}): they
interpreted it as being a broad iron line.

As we show in the next sub-section, the exact form of the soft
components is poorly determined by the \RXTE\ data.
Although we showed a data point below 2 keV in Figure 4, the PCA response
matrix is currently not reliable below 2 keV, and this data point was
dropped from our fitting.
We therefore do not get a good measurement of the roll-over, and hence 
cannot determine $N_H$ reliably.
Thus for all our spectra it has been fixed at 
$N_H = 5 \times 10^{21}~{\rm cm}^{-2}$, as was 
found in the {\it EXOSAT} observations that more reliably measured the 
lower energy spectrum (\cite{ilo86,men97}).

{\it An extra broad emission feature around $\sim 6.4$ keV is always 
required to get a good fit.}
Here we assume it is an iron line feature, that may have been 
broadened by orbital Doppler motions and/or scattering off a hot medium.
Detailed Compton scattering line profiles have been generated
using our Monte Carlo codes and are presented elsewhere (\cite{bls98}).
Here we simply use a broad Gaussian and an iron edge.
We caution that it is possible that the broad \RXTE\ feature seen in the
sources listed above is due to a currently unidentified systematic flaw 
in the PCA response matrices, although the observations of MCG --5-23-16 
make this unlikely.

Unlike for the {\it Ginga}--OSSE observation of GX 339--4 in 1991 
(\cite{zdz98}) our simplistic continuum fitting of the
1996 data does not require a significant reflection component.
Dove et al. (1998) found that no reflection component was needed to 
explain their \RXTE\ observations of Cygnus X-1.
They noted that the broad spectral range covered by \RXTE\ is an
important improvement over previous observations for accurately 
modeling the continuum.
However, we caution that we are only using simple phenomenological models 
in this paper; for a more self-consistent modeling of this data, see
B\"ottcher et al. (1998), which also concludes that there is no need for
a strong reflection component because most of the incident flux from the 
corona goes into heating of the disk surface layer and is not reflected.

Figure 5 shows an example of a complete fit to the \RXTE\ data where we have 
used a simple power law for the soft component.
The PLE component has $\alpha = 1.22 \pm 0.01$, consistent with the OSSE 
best fit value.
The iron emission line is very broad.
Its equivalent width (EW) is 640 eV.

\subsection{Sunyaev-Titarchuk Function Fit}

A ST model can also be used to give a good fit to the \RXTE\ data {\it alone}.
(In the next section we show that {\it the ST model does not give a good fit
to the combined \RXTE-OSSE data.})

Figures 6--8 show three sample fits.
We show three different models for the soft component (power law, black body,
and thermal bremsstrahlung) to illustrate how this affects the relative 
contributions of the soft components.
In all cases, a broad iron emission line is required.
Its EW = 475, 700, and 570 eV in Figures 6--8 respectively.
Note that in the unfolded spectra shown, we have not normalized
the PCA and HEXTE spectra: this is discussed in the next section.

It is important to note that the spectra are very hard.
{\it The ST component dominates the entire \RXTE\ spectrum, even down 
to 2 keV.}
(The PLE and soft power law components cross at 4 keV in Figure 5.) 
This is particularly important in our variability study using this data
(\cite{smi99II}) where we divide the PCA data into three bands
2--5, 5--10, and 10--40 keV to try to highlight the separate components.
There we show that it is the {\it soft 2--5 keV band that is the most
variable.}

\section{Joint \RXTE\--OSSE Spectra}

We now combine the \RXTE\ and OSSE data to generate the joint spectrum.
The results presented in this section should be treated with care, because
the two data sets are not quite simultaneous.
The fact that the OSSE hardness ratio did not change significantly over our
two week observation makes it reasonable to assume that the average gamma-ray 
spectral shape was the same at the time of the \RXTE\ observation, but this 
remains an assumption.

In our rapid variability study of this RXTE data we show that the hardness 
ratios do not change dramatically with time or brightness (\cite{smi99II}).
Thus while the source is extremely variable, with the 2--5 keV band showing
the greatest variability, the average spectra are
representative of the spectral shape throughout the \RXTE\ observation.
{\it However, it should be remembered that the spectrum presented here is
a representative average, and the actual spectrum at any given instant will
have an overall normalization that can differ by a factor $\sim 3$.}

\subsection{Power Law Times Exponential Fit}

The PLE best fits to the separate OSSE (Figure 3(a)) and \RXTE\ (Figure 5) 
data used the same $kT = 96.9$ keV and had consistent values of $\alpha$.
We therefore will get a good joint \RXTE--OSSE fit provided the relative
normalizations between the different instruments can be matched.
This depends on the absolute calibrations of the OSSE, PCA, and HEXTE
instruments.
We could simply shift the unfolded OSSE, PCA, and HEXTE spectra arbitrarily
to make them match.
But instead we have used the \RXTE\ ASM and BATSE data to try to be more 
rigorous.

The uncorrected PCA data gives a 2--10 keV flux of 77 mCrab.
However, the ASM gave a 2--10 keV flux of only 60 mCrab for this day.
This suggests that the unfolded PCA spectrum needs to be normalized by
$\times 0.78$ to get the correct value.

The uncorrected HEXTE data gives a 20--100 keV flux of 
$3.6 \times 10^{-2}~{\rm photons}~{\rm cm}^{-2}{\rm s}^{-1}$.
This is close to the BATSE measured flux of
$4.0 \times 10^{-2}~{\rm photons}~{\rm cm}^{-2}{\rm s}^{-1}$.
This suggests that the unfolded HEXTE spectrum needs to be normalized by
$\times 1.1$ to get the correct value.

We therefore infer that the relative normalization of the HEXTE and PCA
spectra should be 0.71, which is exactly what was found independently 
in the joint fitting.

Based on the 50--70 keV OSSE daily light curve in Figure 2, we would expect 
to have to multiply our average OSSE spectrum by a factor of $(110/94) = 1.17$ 
to get the correct normalization on 1996 July 26.
This agrees with doing a joint \RXTE\--OSSE fit in XSPEC, where we get a 
relative normalization between OSSE and the PCA of 1.5 (which is 1.17/0.78).

Figure 9 shows the effect of shifting the unfolded spectra from Figures 3(a)
and 5 by these amounts.
As expected, the data sets join smoothly.
The model curve is the same as in Figure 5, but with an overall normalizing
factor of $\times 0.78$ that from before (the RXTE fits assumed the PCA was
perfectly calibrated).

Figure 9 highlights that it is extremely important that future observations
of GX 339--4 should make simultaneous X-ray and gamma-ray observations to
accurately measure the spectrum both above and below the cut-off energy.

\subsection{Sunyaev-Titarchuk Function Fit}

The ST best fits to the separate OSSE (Figure 3(b)) and \RXTE\ (Figures 6--8) 
data do not give consistent ST fit parameters.
We therefore do not expect to get a good joint \RXTE--OSSE fit.

Figure 10 shows the results of adding the OSSE data to Figure 6.
We have fixed all the model parameters to those given in Figure 6, except
for the relative normalization of the OSSE and \RXTE\ instruments.
It is clear that the model fit to the \RXTE\ data does not explain the 
shape of the joint \RXTE\--OSSE spectrum: the model 
drops off too rapidly with increasing energies to give an acceptable joint fit.
This now agrees with Grabelsky et al. (1995).

In \S 5 we showed that an ST model could fit the OSSE data alone.
However, we find that these model parameters do not give a good fit 
to the RXTE data.
Even if we ignore all the soft components and just try to use a ST model alone
to simultaneously fit the OSSE data and the \RXTE\ data above 15 keV,
we are unable to get an acceptable fit (the best \chisqnu = 1.5, 
$Q = 4 \times 10^{-9}$, for $kT = 35$ keV, and $\tau = 3.8$).

It is clear that the correct shape cannot be generated by the ST model when 
a larger energy range is available.
We cannot rule out that the gamma-ray spectrum changed dramatically between
the end of our OSSE observation and our \RXTE\ observation, though the 
prior evolution of the source makes this unlikely.
Again this highlights the need for future observations of GX 339--4 to 
make simultaneous X-ray and gamma-ray observations.

More realistic and self-consistent Compton scattering models can explain the 
harder spectrum presented here.
A full study is beyond the scope of this paper, but in 
B\"ottcher, Liang, \& Smith (1998) we develop such a detailed simulation, 
and illustrate that it can fit the joint GX 339--4 spectrum.

\section{Summary}

As part of our multiwavelength campaign of observations of GX 339--4 in 1996
we presented our radio, X-ray, and gamma-ray observations made in July, when
the source was in a hard state (= soft X-ray low state).

The radio observations were made at the time when there was a possible 
radio jet.
We showed that the radio spectrum was flat and significantly variable, and
that the radio spectral shape and amplitude were not anomalous for the 
source at this time.
Daily light curves from our pointed OSSE observation July 9--23,
from BATSE, and from the ASM on \RXTE\ also showed that there was no 
significant change in the X- and gamma-ray flux or hardness during the 
time the radio jet-like feature was seen.

The higher energy portion of our pointed \RXTE\ observation made July 26 
is equally well fit using simple PLE and ST functions.
An additional soft component is required, as well as a broad emission
feature centered on $\sim 6.4$ keV.
This may be an iron line that is broadened by orbital Doppler motions and/or
scattering off a hot medium.
Its equivalent width is $\sim 600$ eV.
Both a PLE and a ST model also fit our OSSE spectrum on its own.
Although the observations are not quite simultaneous, combining the \RXTE\ 
and \CGRO\ spectra we find that the PLE model easily fits the joint spectrum.
However, the ST model 
drops off too rapidly with increasing energies to give an acceptable joint fit.

Our results show that it is extremely important that future studies of 
GX 339--4 should make truly simultaneous multiwavelength observations,
particularly given the variability of the source.
It is essential to accurately measure the spectrum both above and below 
the gamma-ray cut-off energy.
It will also be extremely interesting to combine future unusual radio
activity with the behavior at high energies.

\acknowledgments

We thank the referee for carefully reading the manuscript and providing
useful suggestions and clarifications.
This work was supported by NASA grants NAG 5-1547 and 5-3824 at 
Rice University.
This work made use of the \RXTE\ ASM data products provided by the ASM/\RXTE\ 
teams at MIT and at the \RXTE\ SOF and GOF at NASA's Goddard Space Flight 
Center.
The BATSE daily light curves were provided by the Compton
Observatory Science Support Center at NASA's Goddard Space Flight Center.

\clearpage

\begin{figure} \plotone{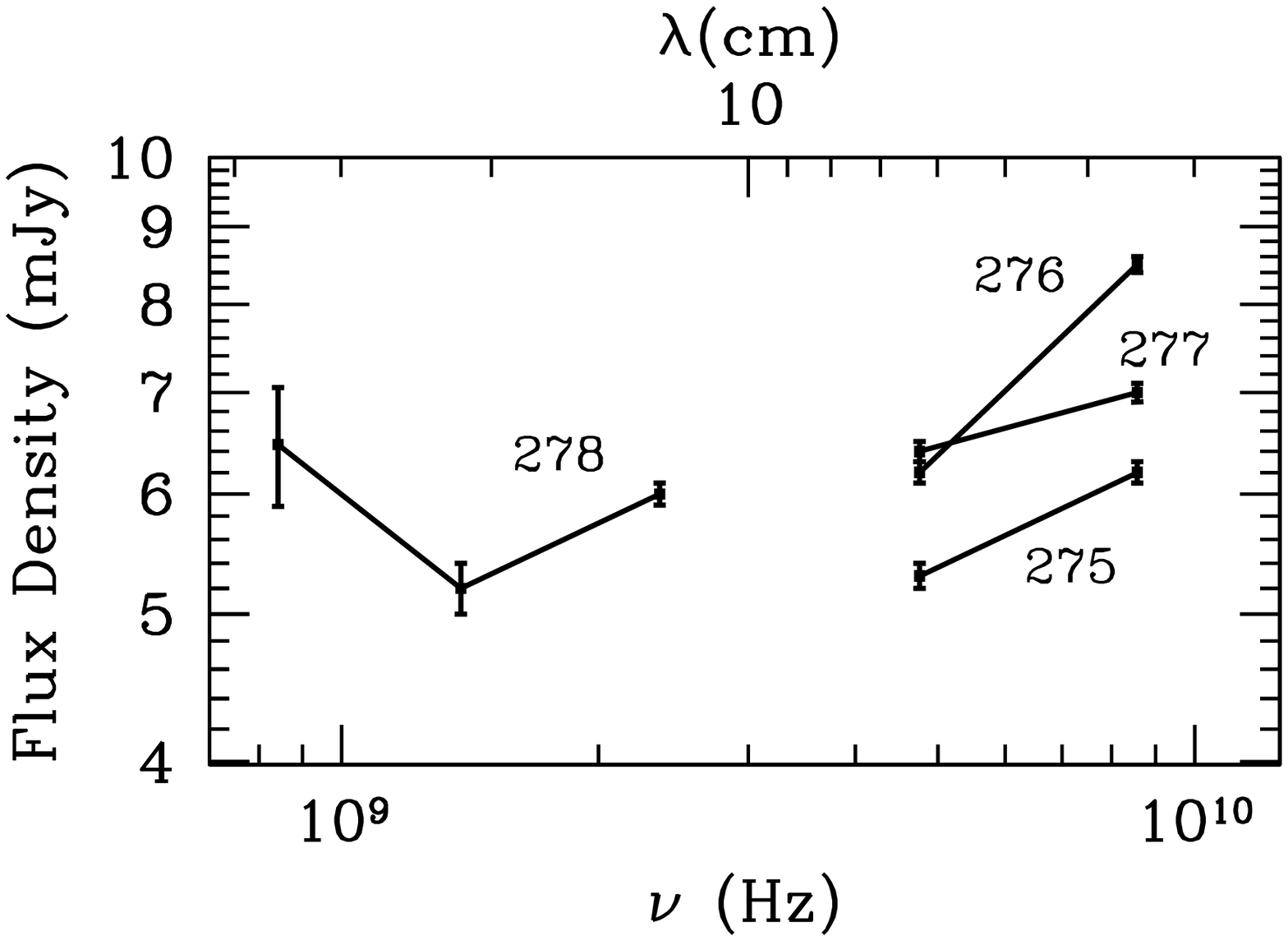} \begin{center}
\figcaption{Daily radio spectra in 1996 July.
All the data points are from ATCA, except the one at 843 MHz from MOST.
Labels are MJD - 50000: 1996 July 11 is MJD 50275.}
\end{center} \end{figure}

\begin{figure} \plotone{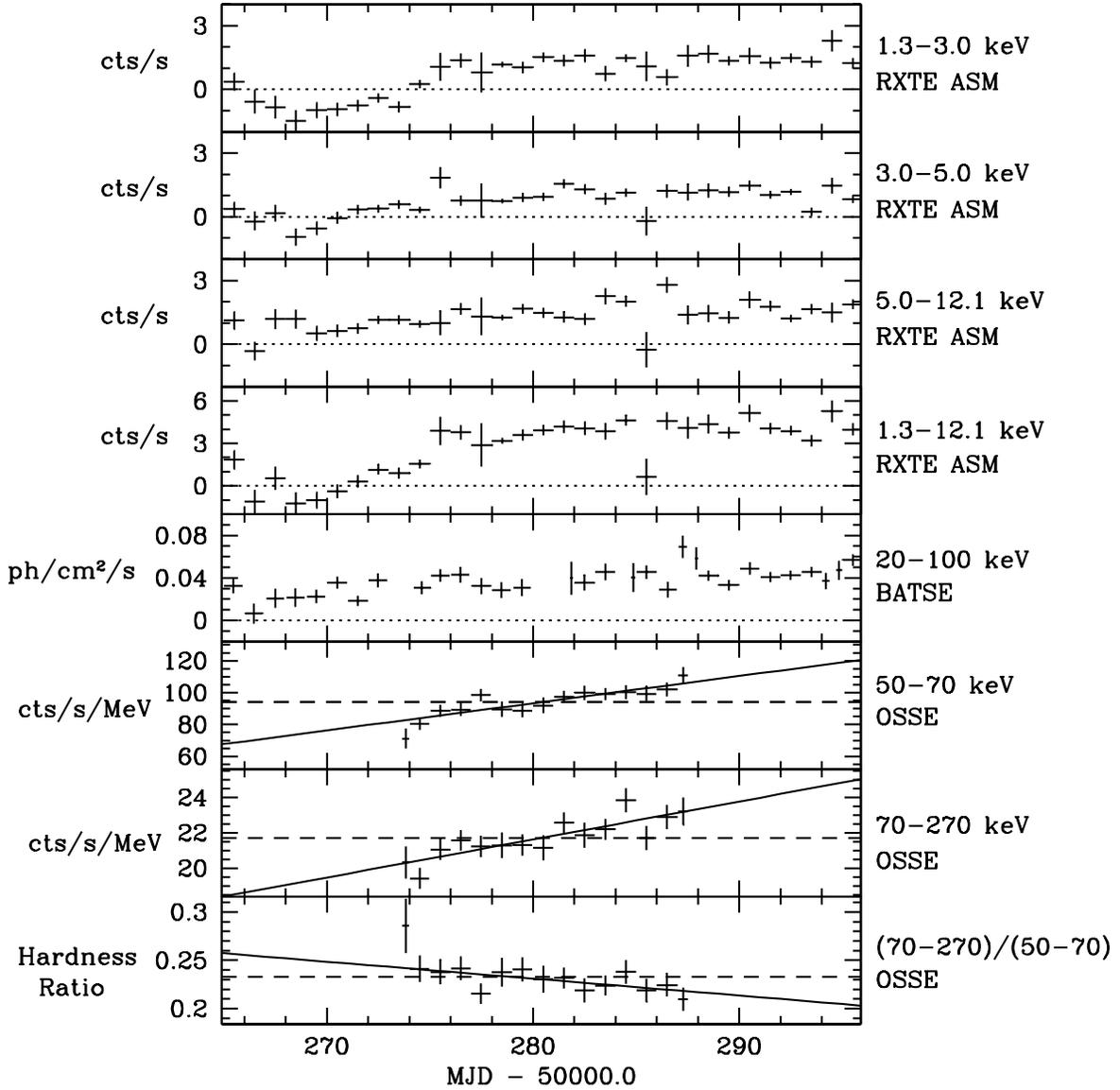} \begin{center}
\figcaption{Daily light curves for 1996 July 1--31.
From top to bottom, the panels are:
\RXTE\ ASM 1.3--3.0, 3.0--5.0, 5.0--12.1, and 1.3--12.1 keV,
BATSE Earth occultation 20--100 keV (assuming a fixed OTTB model with 
$kT = 60$ keV),
OSSE 50--70, and 70--270 keV, and
OSSE hardness ratio (70--270 keV)/(50--70 keV).
The dashed lines show the best constant fits to the OSSE data.
The solid lines show weighted least squares linear fits to the OSSE data.
The possible radio jet-like feature was seen MJD 50275--7.}
\end{center} \end{figure}

\begin{figure} \plottwo{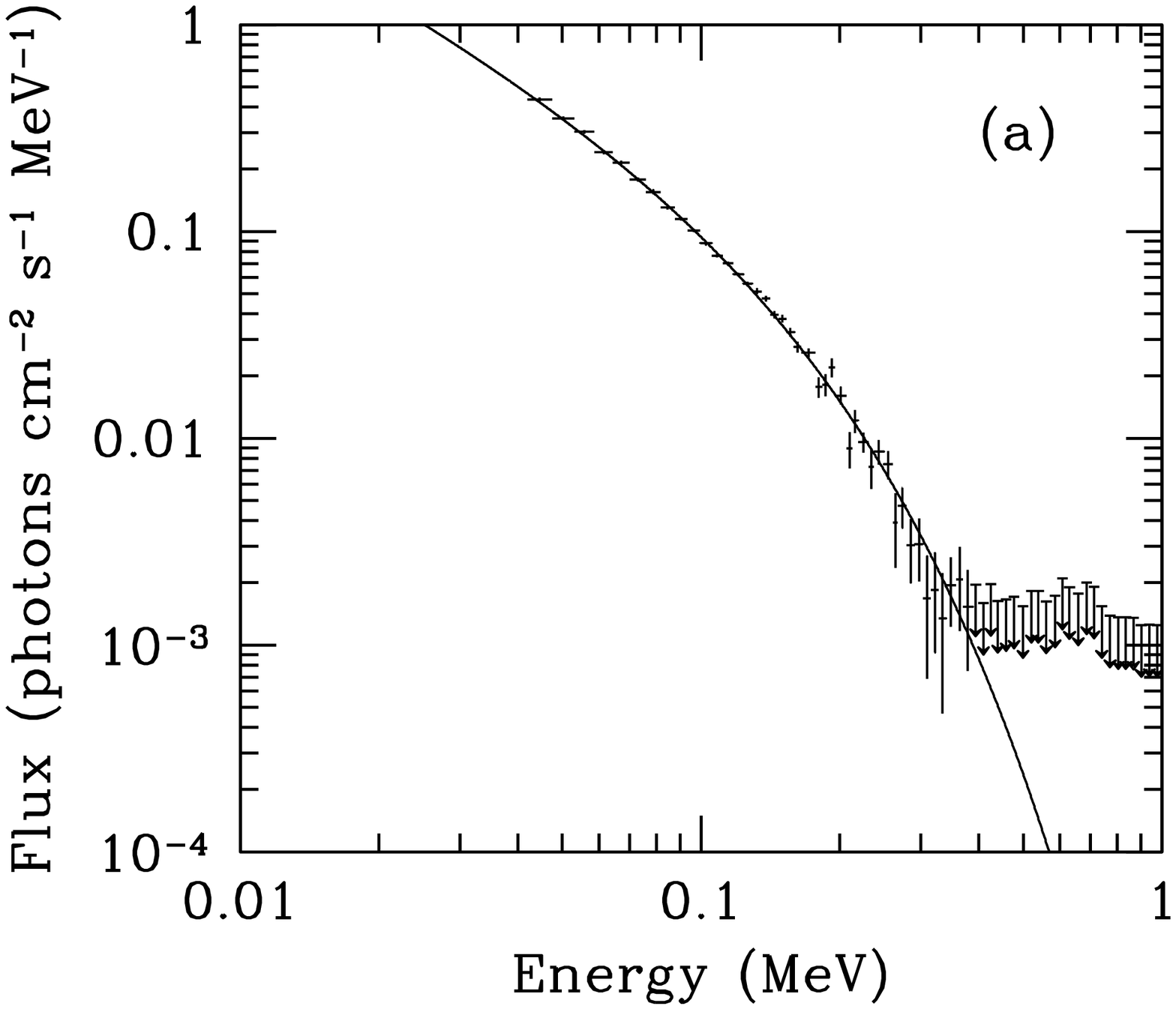}{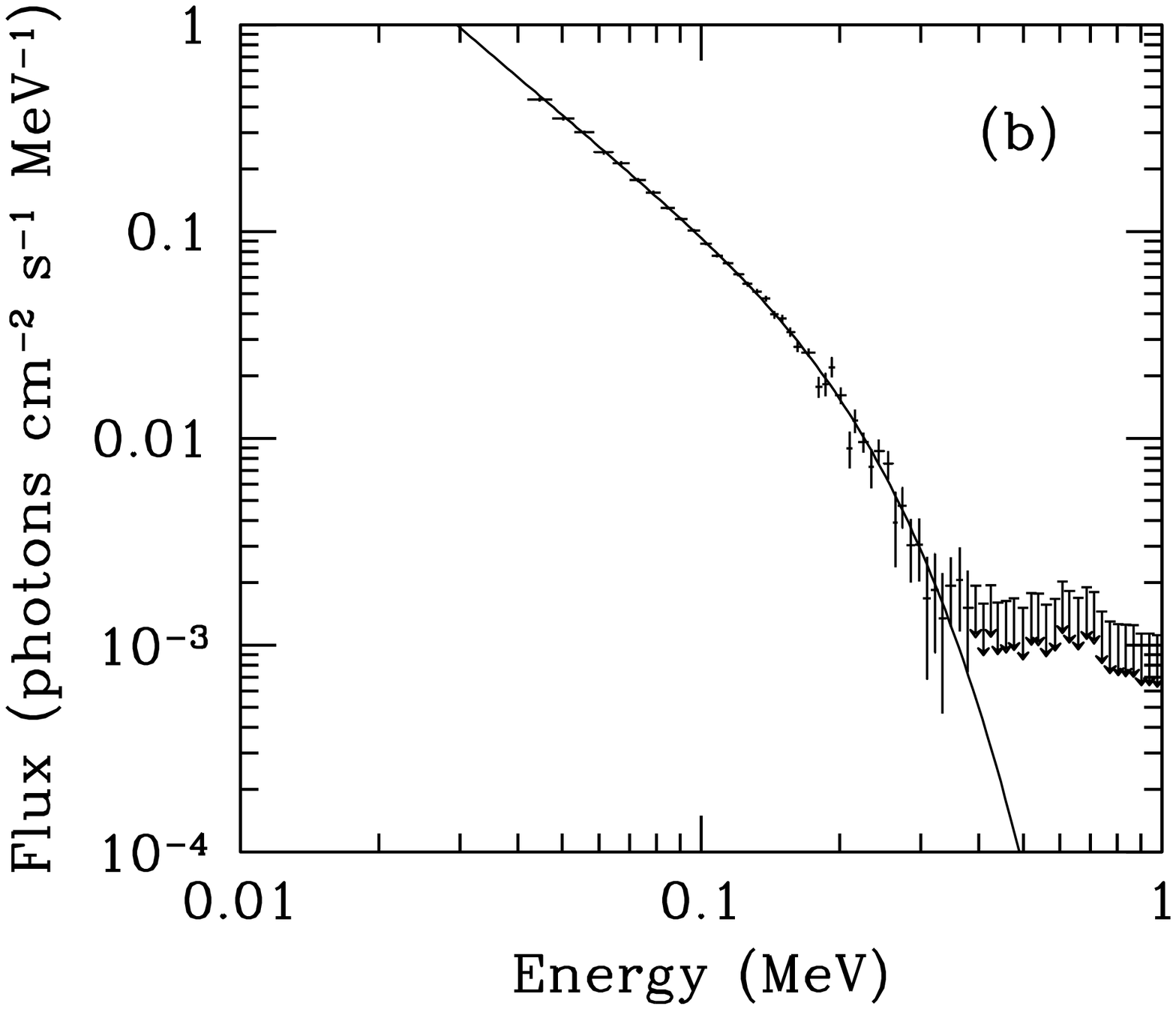} \begin{center}
\figcaption{1996 July 9--23 OSSE spectrum of GX 339--4 integrated over the 
whole two week observation.
(a) best fit PLE model to OSSE data alone.
(b) best fit ST model to OSSE data alone.
For easy comparison with Grabelsky et al. (1995), we use MeV units on
both axes: all the other spectra in this paper use keV units.
The upper limits are $2 \sigma$.}
\end{center} \end{figure}

\begin{figure} \plotone{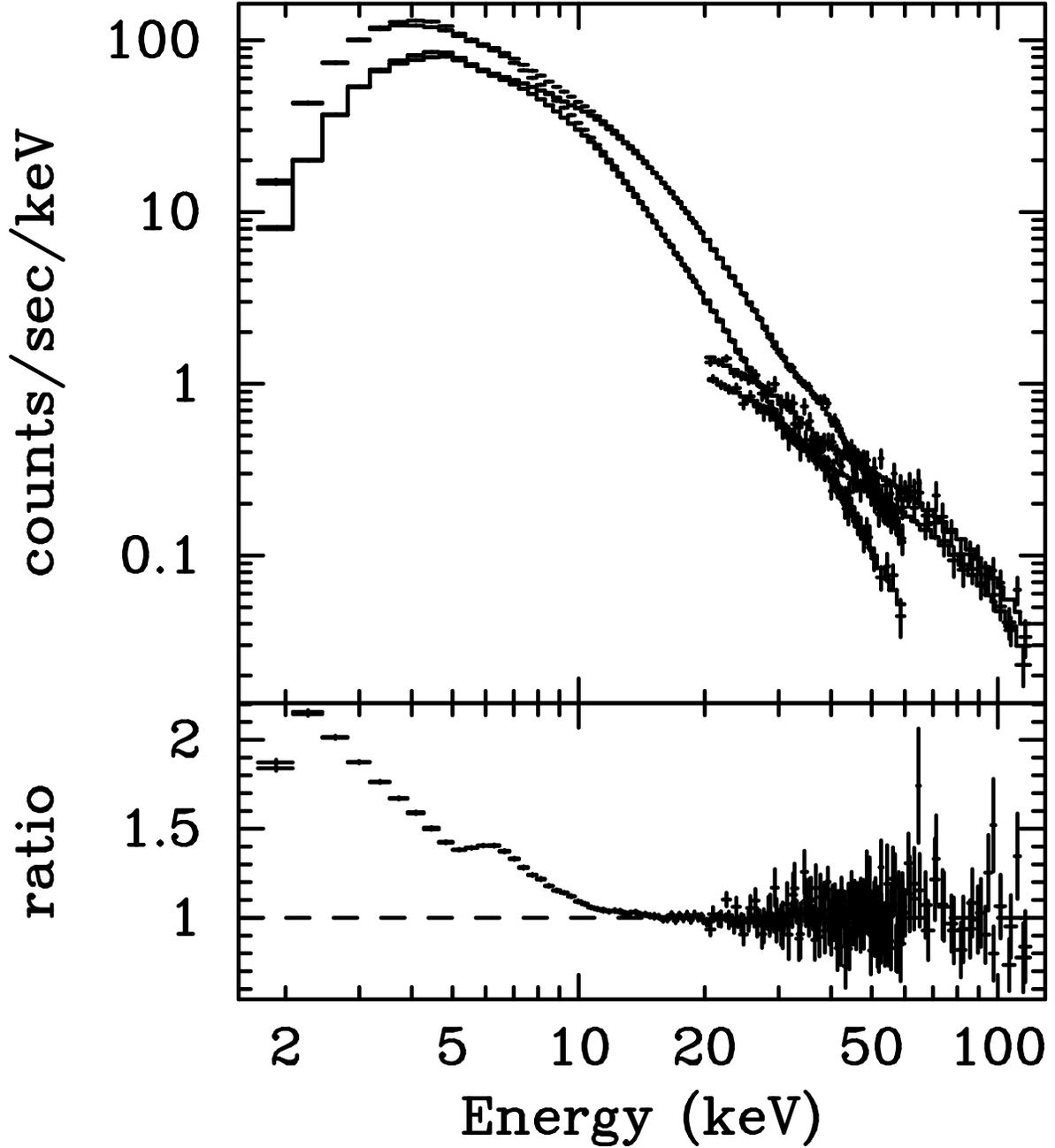} \begin{center}
\figcaption{1996 July 26 \RXTE\ observation of GX 339--4.
PLE fit to data above 15 keV only.
In all the \RXTE\ figures, the top PCA curve (2 -- 60 keV) uses all PCU, the 
bottom PCA curve uses just the top PCU, the top HEXTE curve (20 -- 120 keV)
is from cluster A, and the bottom HEXTE curve is from cluster B.  
The bottom panel plots the ratio of the data to the PLE model.
Note that since this figure is for illustrative purposes, data below
2 keV and around the Xenon L edge are included: these are not used in
our final fitting.}
\end{center} \end{figure}

\begin{figure} \plotone{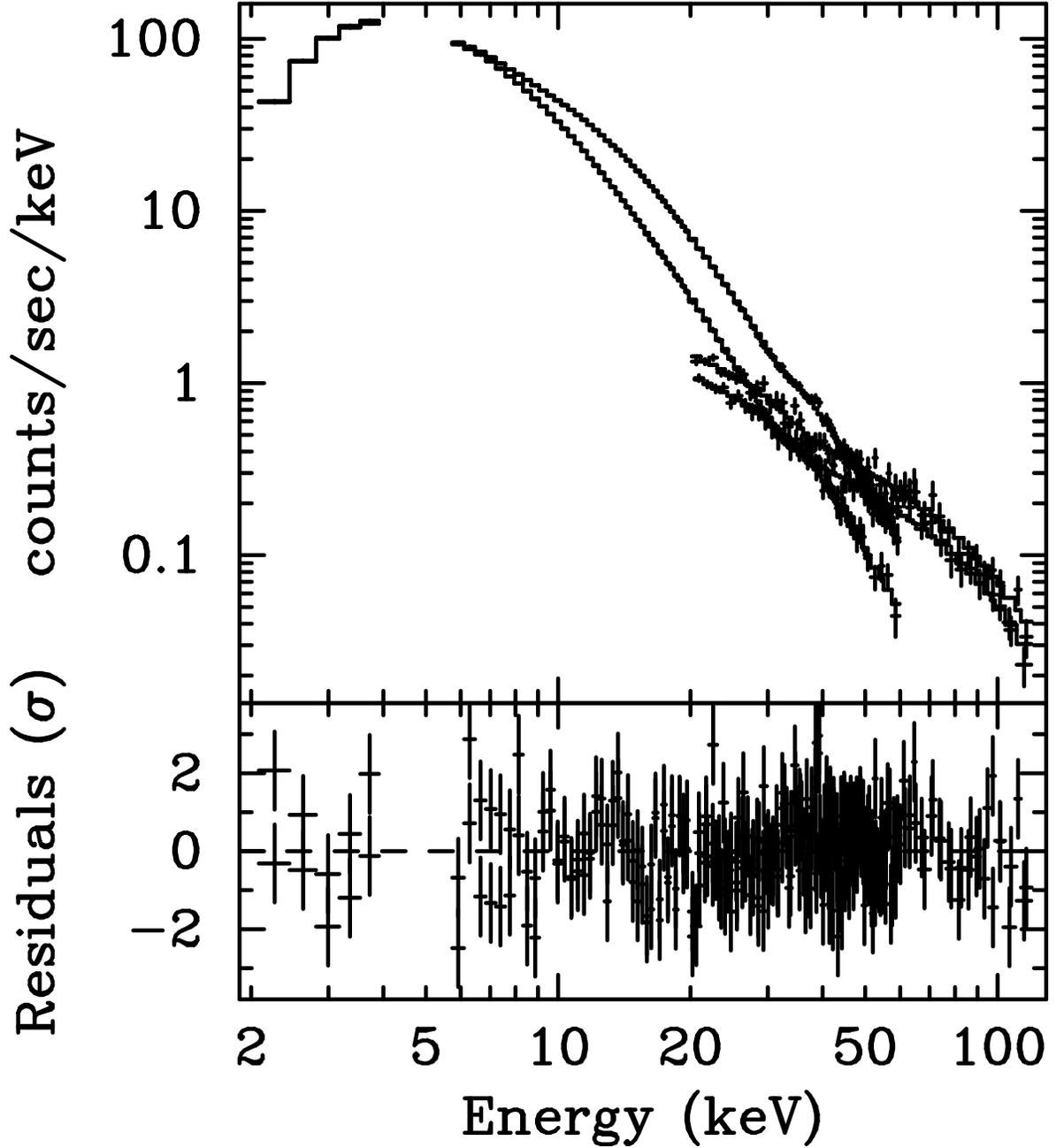} \begin{center}  \vspace{-0.25in}
\figcaption{\small{PLE model fit to 1996 July 26 \RXTE\ observation of 
GX 339--4.
The PLE component has $\alpha = 1.22 \pm 0.01$, $kT = 96.9$ keV (fixed), and 
flux $0.111 \pm 0.005~{\rm photons}~{\rm cm}^{-2}{\rm s}^{-1}{\rm keV}^{-1}$
at 1 keV.
The soft component is a power law with photon index $3.00 \pm 0.06$ and
flux $1.01 \pm 0.05~{\rm photons}~{\rm cm}^{-2}{\rm s}^{-1}{\rm keV}^{-1}$
at 1 keV.
$N_H = 5 \times 10^{21}~{\rm cm}^{-2}$ (fixed).
The Gaussian line has centroid $6.6 \pm 0.3$ keV, 
width $\sigma = 1.7 \pm 0.1$ keV, and a total of 
$0.009 \pm 0.001~{\rm photons}~{\rm cm}^{-2}{\rm s}^{-1}$ in the line.
The edge has threshold energy $7.18 \pm 0.08$ keV and absorption depth
$0.09 \pm 0.02$ at the threshold.
The relative overall normalization of the HEXTE and PCA spectra 
is $0.71 \pm 0.01$.
The errors are 90\% confidence regions for varying one parameter.
This fit has \chisqnu = 1.025, $Q = 0.36$.
The residuals are given in terms of sigmas, with error bars of size one.}}
\end{center} \end{figure}

\begin{figure} \plotone{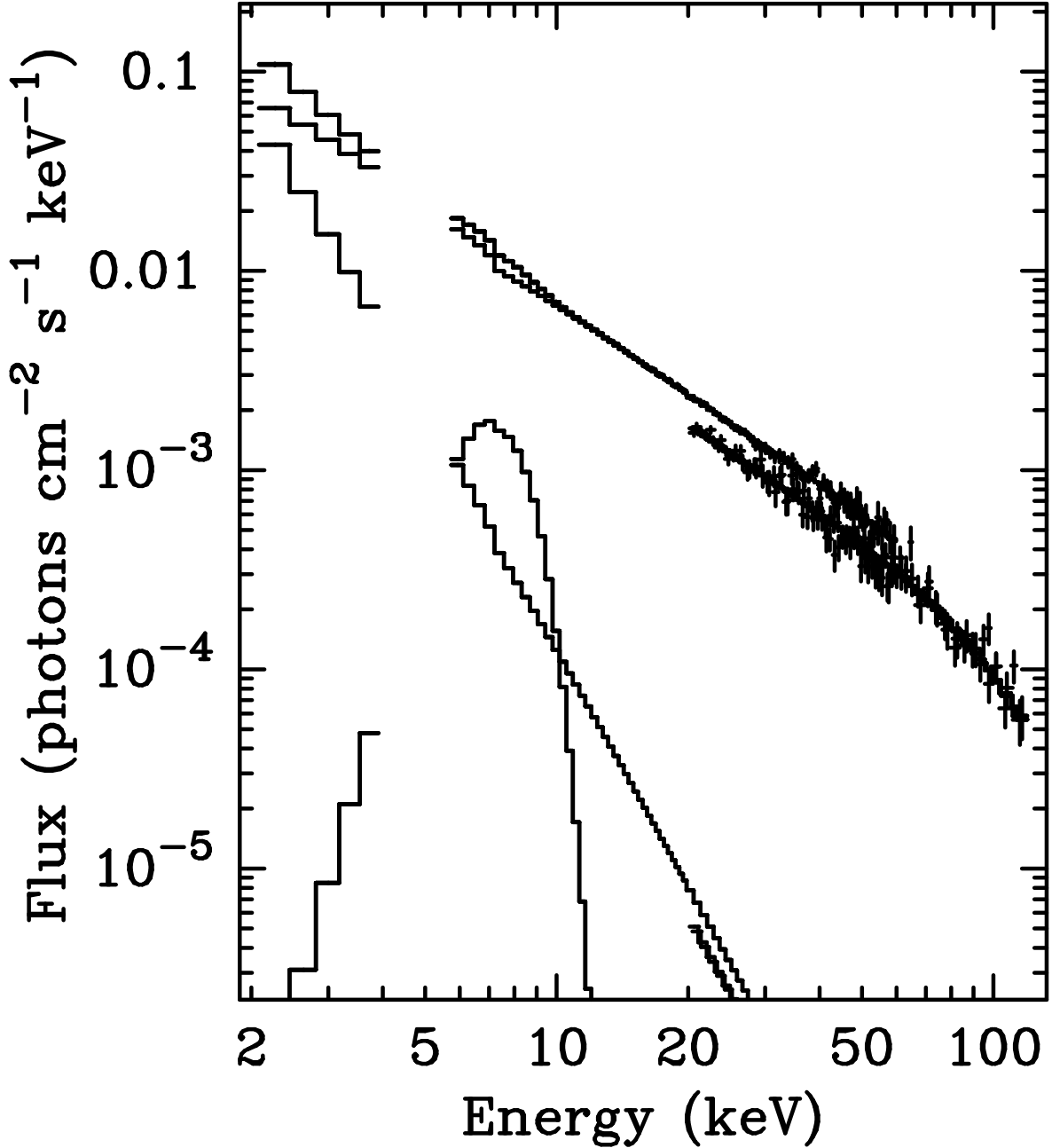} \begin{center} \vspace{-0.25in}
\figcaption{\small{Example of ST model fit to 1996 July 26 \RXTE\ 
observation of GX 339--4.
The top curve is the total unfolded model spectrum (not normalized), 
and the other curves show the hard, soft, and iron line components.
The ST component has $\tau = 4.9 \pm 0.2$ and $k T = 23 \pm 1$ keV.
The ST normalization $\kappa = 0.30 \pm 0.01$ is given by $N f / 4 \pi d^2$, 
where $N$ is the total number of photons from the source, $d$ is the distance 
to the source, $f = z(z+3)y^z/\Gamma(2z+4)/\Gamma(z)$, $z$ is the spectral 
index, $y$ the injected photon energy in units of the temperature, and
$\Gamma$ is the incomplete gamma function.
The soft component is a power law with photon index $4.0 \pm 0.1$ and
flux $1.3 \pm 0.1~{\rm photons}~{\rm cm}^{-2}{\rm s}^{-1}{\rm keV}^{-1}$
at 1 keV.
$N_H = 5 \times 10^{21}~{\rm cm}^{-2}$ (fixed).
The Gaussian line has centroid $7.2 \pm 0.2$ keV, 
width $\sigma = 1.3 \pm 0.1$ keV, and a total of 
$0.0059 \pm 0.0003~{\rm photons}~{\rm cm}^{-2}{\rm s}^{-1}$ in the line.
The edge has threshold energy $7.12 \pm 0.06$ keV and absorption depth
$0.14 \pm 0.02$ at the threshold.
The relative overall normalization of the HEXTE and PCA spectra 
is $0.71 \pm 0.01$.
The errors are 90\% confidence regions for varying one parameter.
This fit has \chisqnu = 0.99, $Q = 0.54$.}}
\end{center} \end{figure}

\begin{figure} \plotone{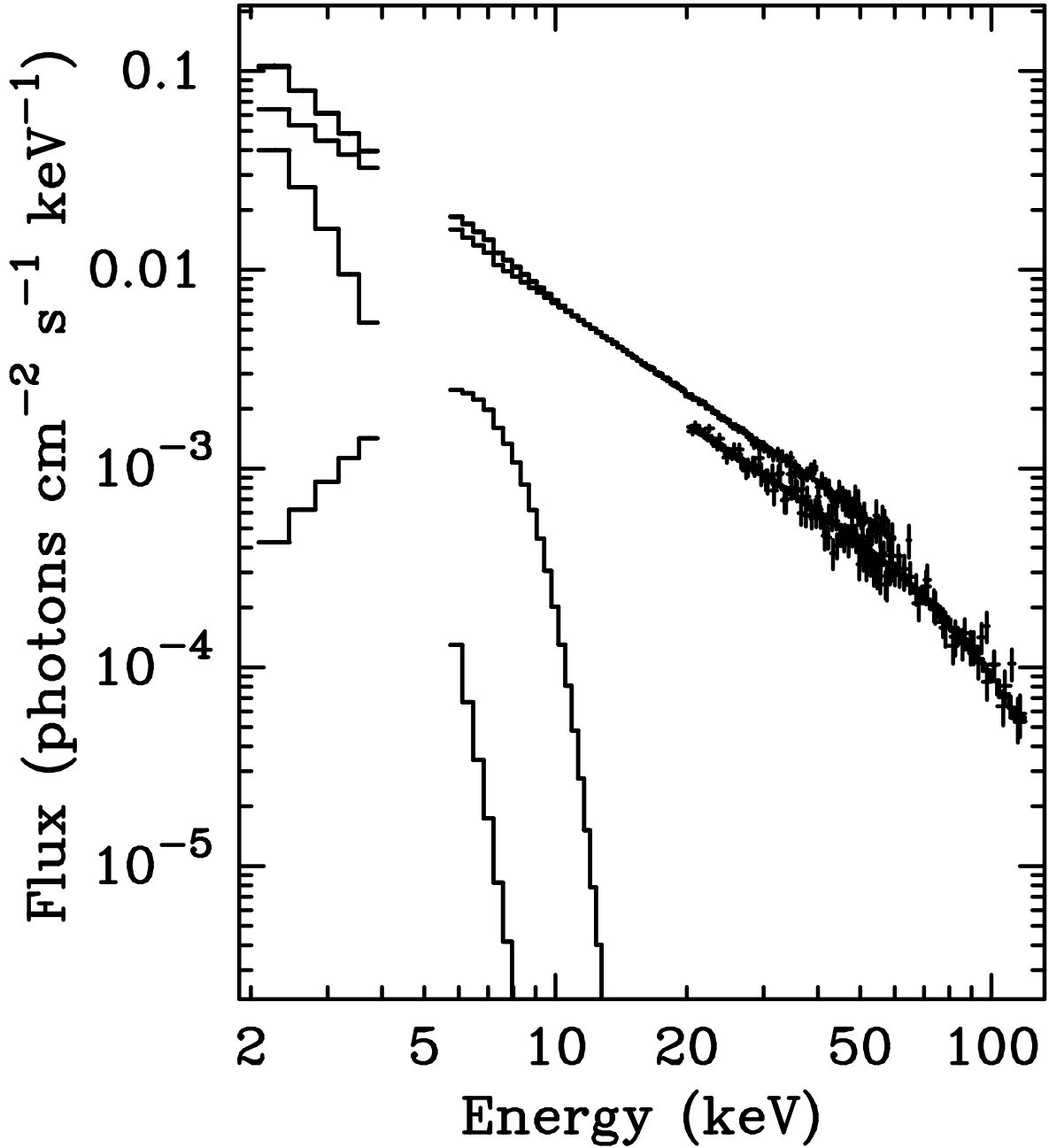} \begin{center} \vspace{-0.25in}
\figcaption{\small{Example of ST model fit to 1996 July 26 \RXTE\ 
observation of GX 339--4.
The top curve is the total unfolded model spectrum (not normalized), 
and the other curves show the hard, soft, and iron line components.
The ST component has $\tau = 5.0 \pm 0.2$, $k T = 23 \pm 1$ keV,
and normalization $\kappa = 0.29 \pm 0.01$.
The soft component is a black body with $kT = 0.47 \pm 0.02$ keV and
normalization $L_{39}/D^2_{10} = 0.0069 \pm 0.0005$, where 
$L_{39}$ is the black body luminosity in units of 
$10^{39}$ ergs/sec and $D_{10}$ is the source distance in units of 10 kpc.
Assuming a distance of 4 kpc gives $L = 1.1 \times 10^{36}$ ergs/sec in
the soft component.
$N_H = 5 \times 10^{21}~{\rm cm}^{-2}$ (fixed).
The Gaussian line has centroid $5.7 \pm 0.7$ keV, 
width $\sigma = 1.9 \pm 0.3$ keV, and a total of 
$0.012 \pm 0.003~{\rm photons}~{\rm cm}^{-2}{\rm s}^{-1}$ in the line.
The edge has threshold energy $7.23 \pm 0.15$ keV and absorption depth
$0.06 \pm 0.03$ at the threshold.
The relative overall normalization of the HEXTE and PCA spectra 
is $0.71 \pm 0.01$.
The errors are 90\% confidence regions for varying one parameter.
This fit has \chisqnu = 1.19, $Q = 0.01$.}}
\end{center} \end{figure}

\begin{figure} \plotone{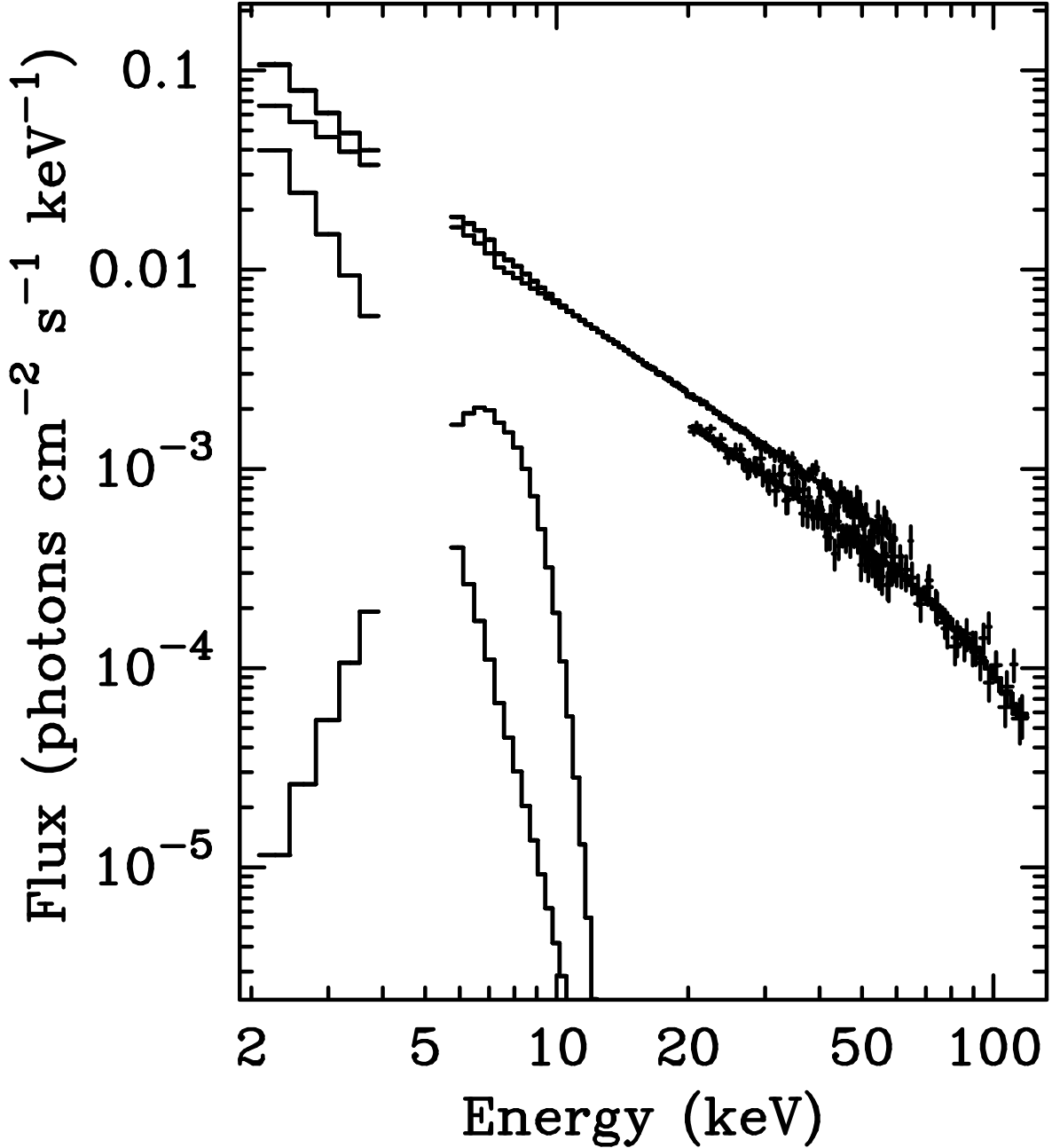} \begin{center} \vspace{-0.25in}
\figcaption{\small{Example of ST model fit to 1996 July 26 \RXTE\ 
observation of GX 339--4.
The top curve is the total unfolded model spectrum (not normalized), 
and the other curves show the hard, soft, and iron line components.
The ST component has $\tau = 4.82 \pm 0.15$, $k T = 23.6 \pm 1.2$ keV,
and normalization $\kappa = 0.303 \pm 0.008$.
The soft component is a thermal bremsstrahlung with $kT = 1.07 \pm 0.04$ keV 
and normalization 
$(3.02 \times 10^{-15}/(4 \pi D^2)) \int n_e n_I dV = 0.88 \pm 0.06$
where $D$ is the source distance (in cm) and $n_e$ and $n_I$ are the
electron and ion densities (in cm$^{-3}$).
$N_H = 5 \times 10^{21}~{\rm cm}^{-2}$ (fixed).
The Gaussian line has centroid $6.8 \pm 0.3$ keV, 
width $\sigma = 1.4 \pm 0.1$ keV, and a total of 
$0.0075 \pm 0.0006~{\rm photons}~{\rm cm}^{-2}{\rm s}^{-1}$ in the line.
The edge has threshold energy $7.10 \pm 0.07$ keV and absorption depth
$0.12 \pm 0.03$ at the threshold.
The relative overall normalization of the HEXTE and PCA spectra 
is $0.71 \pm 0.01$.
The errors are 90\% confidence regions for varying one parameter.
This fit has \chisqnu = 1.03, $Q = 0.31$.}}
\end{center} \end{figure}

\begin{figure} \plotone{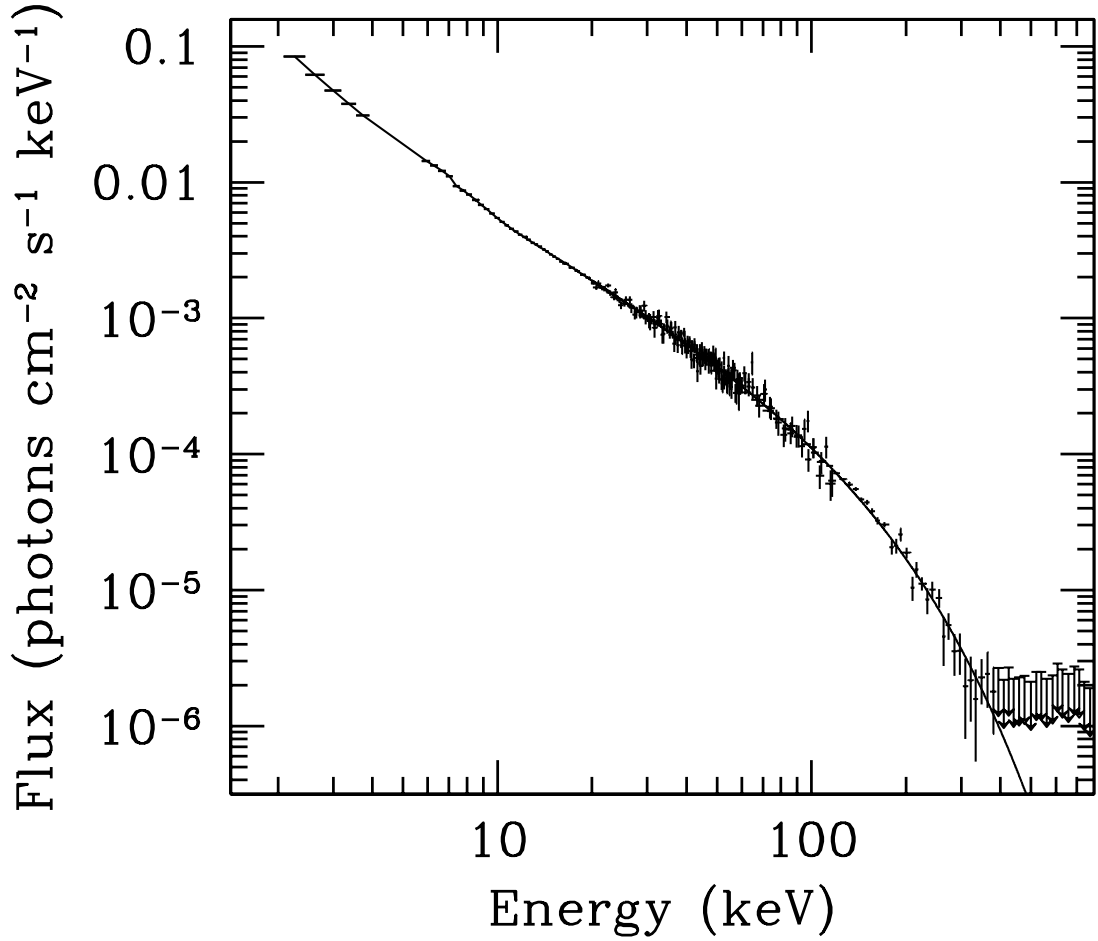} \begin{center}
\figcaption{Joint \RXTE--OSSE spectrum and PLE model fit for GX 339--4.
Note that the two data sets are not quite simultaneous, and the
source is highly variable, so this figure should be used with care.
The PCA data from Figure 5 has been scaled by $\times 0.78$.
The HEXTE data from Figure 5 has been scaled by $\times 1.1$.
The OSSE data from Figure 3(a) has been scaled by $\times 1.17$.
The model fit from Figure 5 has been scaled by $\times 0.78$.}
\end{center} \end{figure}

\begin{figure} \plotone{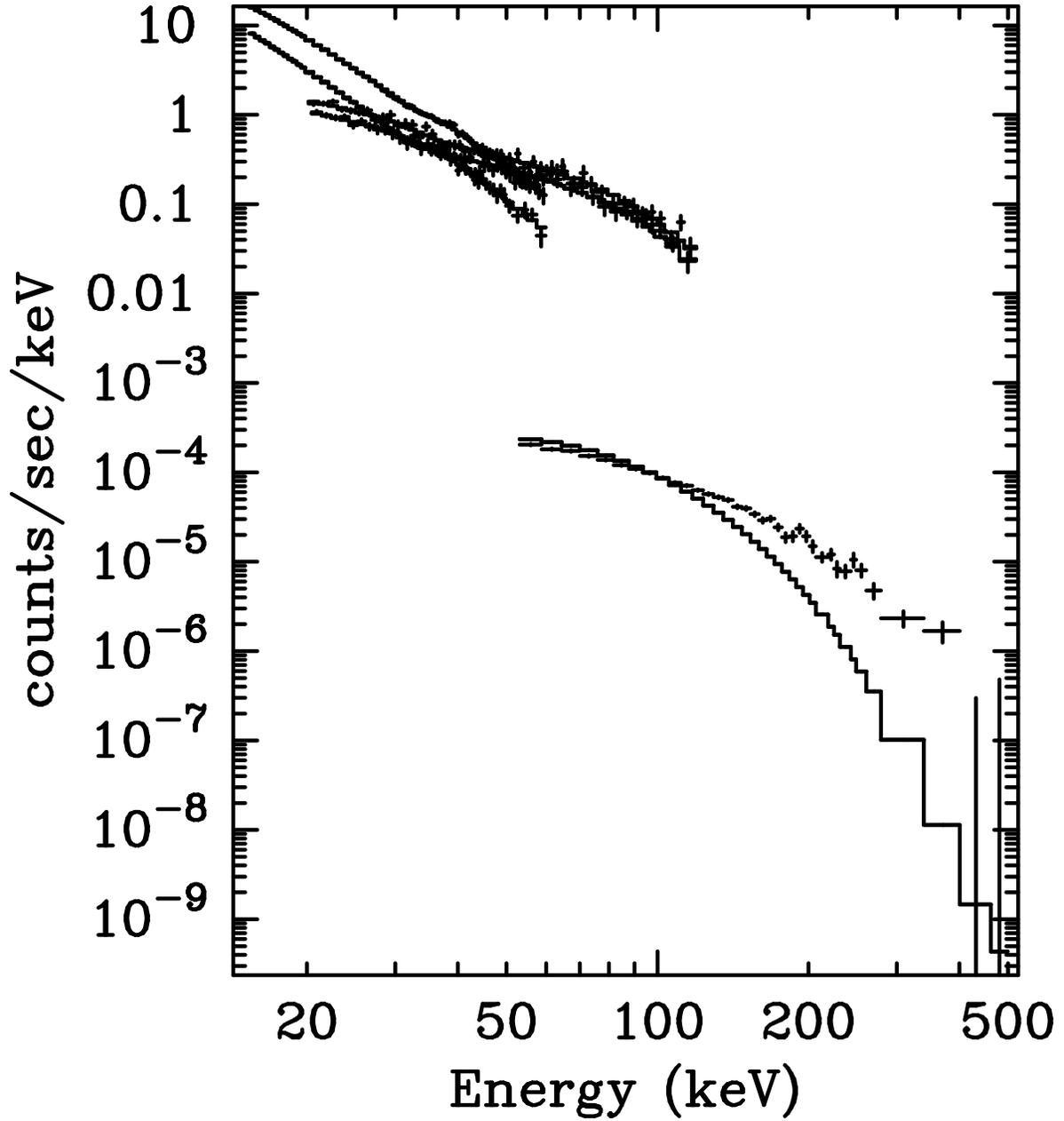} \begin{center}
\figcaption{Joint \RXTE--OSSE spectrum and ST model fit for GX 339--4.
The OSSE data has been added to Figure 6.
All the model parameters are the same as in Figure 6, with the relative 
normalization of the OSSE and \RXTE\ instruments the only parameter that
has been fit here.}
\end{center} \end{figure}

\end{document}